\documentclass[twocolumn,showpacs,aps,prd,floatfix]{revtex4}
\usepackage{graphicx}
\usepackage{dcolumn}
\usepackage{epsfig}
\usepackage{psfrag}
\usepackage{amsmath}
\usepackage{longtable}
\usepackage{bm}
\usepackage{relsize}
\usepackage{xspace}
\long\def\inst#1{\par\nobreak\kern 4pt\nobreak
    {\it #1}\par\vskip 10pt plus 3pt minus 3pt}
\def\babar{\mbox{\slshape B\kern-0.1em{\smaller A}\kern-0.1em
    B\kern-0.1em{\smaller A\kern-0.2em R}}}
\def\pep2{PEP-II}
\def\Kp    {\ensuremath{K^+}\xspace}
\def\Km    {\ensuremath{K^-}\xspace}
\def\piz   {\ensuremath{\pi^0}\xspace}
\def\pip   {\ensuremath{\pi^+}\xspace}
\def\pim   {\ensuremath{\pi^-}\xspace}
\def\Dz{D^0}
\def\KS    {\ensuremath{K^0_{\scriptscriptstyle S}}\xspace}
\def\D       {\ensuremath{D}\xspace}
\def\Dp      {\ensuremath{D^+}\xspace}
\def\Dm      {\ensuremath{D^-}\xspace}
\def\Dz      {\ensuremath{D^0}\xspace}
\def\Dzb     {\ensuremath{\Dbar^0}\xspace}
\def\DzDzb   {\ensuremath{\Dz {\kern -0.16em \Dzb}}\xspace}
\def\DpDm    {\ensuremath{\Dp {\kern -0.16em \Dm}}\xspace}
\def\Dstar   {\ensuremath{D^*}\xspace}
\def\Dstarb  {\ensuremath{\Dbar^*}\xspace}
\def\Dstarz  {\ensuremath{D^{*0}}\xspace}

\def\Dstarp  {\ensuremath{D^{*+}}\xspace}

\def\Ds      {\ensuremath{D^+_s}\xspace}
\def\Dsa {\ensuremath{D^*_{s1}(2710)^+}\xspace}
\def\Dsb {\ensuremath{D_{sJ}^{*}(2860)^+}\xspace}
\def\Dsc {\ensuremath{D_{sJ}(3040)^+}\xspace}

\def\invfb   {\ensuremath{\mbox{\,fb}^{-1}}\xspace}
\def\Dbar    {\kern 0.2em\overline{\kern -0.2em D}{}\xspace}
\def\Db      {\ensuremath{\Dbar}\xspace}
\def\DDb     {\ensuremath{\D {\kern -0.16em \Db}}\xspace}
\def\DDbX     {\ensuremath{\D {\kern -0.16em \Db}X}\xspace}

\def\DstarDstarb     {\ensuremath{\Dstar{\kern -0.16em \Dstarb}}\xspace}

\def\Y#1S{\ensuremath{\Upsilon{(#1S)}}\xspace}
\def\FourS {\Y4S}

\def\sys {\hbox{syst}}
\def\sta {\hbox{stat}}

\newcommand{\gevc}{\ensuremath{{\mathrm{\,Ge\kern -0.1em V\!/}c}}\xspace}
\newcommand{\mevc}{\ensuremath{{\mathrm{\,Me\kern -0.1em V\!/}c}}\xspace}
\newcommand{\gevcc}{\ensuremath{{\mathrm{\,Ge\kern -0.1em V\!/}c^2}}\xspace}
\newcommand{\mevcc}{\ensuremath{{\mathrm{\,Me\kern -0.1em V\!/}c^2}}\xspace}
\newcommand{\mev}{\ensuremath{\mathrm{\,Me\kern -0.1em V\!}}\xspace}
\newcommand{\gev}{\ensuremath{\mathrm{\,Ge\kern -0.1em V\!}}\xspace}
\newcommand{\gevcccc}{\ensuremath{{\mathrm{\,Ge\kern -0.1emV^2\!/}c^4}}\xspace}

\begin{document}
\newcommand{\BaBarYear}    {09}
\newcommand{\BaBarNumber}  {024}
\newcommand{\BABARProcNumber} {\phantom{14}}
\newcommand{\SLACPubNumber} {13747}
\newcommand{\LANLNumber} {yymm.nnnn [hep-ex]}
\newcommand{\BaBarType}      {PUB}  
\begin{flushleft}
\babar-\BaBarType-\BaBarYear/\BaBarNumber \\
SLAC-PUB-\SLACPubNumber \\
\end{flushleft}
\title{\boldmath Study of $D_{sJ}$ decays to $D^*K$ in inclusive $e^+e^-$ interactions}

\author{B.~Aubert}
\author{Y.~Karyotakis}
\author{J.~P.~Lees}
\author{V.~Poireau}
\author{E.~Prencipe}
\author{X.~Prudent}
\author{V.~Tisserand}
\affiliation{Laboratoire d'Annecy-le-Vieux de Physique des Particules (LAPP), Universit\'e de Savoie, CNRS/IN2P3,  F-74941 Annecy-Le-Vieux, France}
\author{J.~Garra~Tico}
\author{E.~Grauges}
\affiliation{Universitat de Barcelona, Facultat de Fisica, Departament ECM, E-08028 Barcelona, Spain }
\author{M.~Martinelli$^{ab}$}
\author{A.~Palano$^{ab}$ }
\author{M.~Pappagallo$^{ab}$ }
\affiliation{INFN Sezione di Bari$^{a}$; Dipartimento di Fisica, Universit\`a di Bari$^{b}$, I-70126 Bari, Italy }
% ...
\collaboration{The \babar\ Collaboration}
\noaffiliation

\begin{abstract}
We observe the decays $\Dsa \to D^*K$ and $\Dsb \to D^*K$ and measure their branching fractions relative to the $D K$ final state. We also observe, in the $D^*K$ mass spectrum,
a new broad structure
at a mass of  $(3044 \pm 8_{\sta}\,(^{+30}_{-5})_{\sys})$ \mevcc having a width
$\Gamma=(239 \pm 35_{\sta}\,(^{+46}_{-42})_{\sys})$  \mev.  
To obtain this result we use 470~${\rm fb}^{-1}$ of data
recorded by the \babar\  detector at the \pep2 asymmetric-energy $e^+e^-$
storage rings at the Stanford Linear Accelerator Center 
running at center-of-mass energies near 10.6 GeV.
\end{abstract}


\maketitle
\section{\boldmath Introduction}
The spectrum of known $c\overline{s}$ states can be described as 
two S-wave states ($D_s^+$, $D_s^{*+}$) with $J^{P}=0^-,1^-$, and four P-wave 
states ($D^*_{s0}(2317)^+$, $D_{s1}(2460)^+$, $D_{s1}(2536)^+$, $D_{s2}^*(2573)^+$) with 
$J^P=0^+,1^+,1^+,2^+$. Whether this picture is correct remains controversial 
because the states at 2317 \mevcc and 2460 \mevcc~\cite{dsnew} had been expected to lie at much 
higher masses~\cite{th}. Since the discovery of the new $D_s^+$ mesons much theoretical work has been done,
however new experimental results are needed in order to understand this sector
of spectroscopy.

Recently, two new $\Ds$ mesons have been discovered, $\Dsa$~\cite{babar_ds,belle_ds} and $\Dsb$~\cite{babar_ds}.
The analysis of $\Dsa$ produced in $B$ decays gives the assignment $J^{P}=1^-$. For $\Dsb$ assignments of $J^P=0^+$~\cite{port,close} and $J^P=3^-$~\cite{colangelo1} have been proposed.

We report here on a search for new $D^+_{sJ}$ mesons in the mass spectrum of $D^{(*)}K$ inclusively produced at the \pep2
asymmetric-energy $e^+e^-$ storage rings and 
recorded  by the \babar\ detector.
This paper is organized as follows. In Sect. II we give a short description of the \babar \ experiment and in Sect. III we 
describe the data selection. Section IV is devoted to the study of the $D K$ system
and in Sect. V we present the study of the $D^*K$ system.
In Sect. VI we describe fits to the $D^*K$ mass spectrum while in Sect. VII we present an analysis of the angular distributions.
Measurements of ratios of branching fractions are described in Sect. VIII
and we summarize the results in Sect. IX.
  
\section{\boldmath The \babar\ experiment}
This analysis is based on
 a 470~$\invfb$ data sample recorded at the
\FourS resonance and 40 MeV below the resonance.
The \babar\ detector is
described in detail elsewhere~\cite{babar}. We mention here only the parts of the 
detector which are used in the present analysis.
Charged particles are detected
and their momenta measured with a combination of a 
cylindrical drift chamber (DCH)
and a silicon vertex tracker (SVT), both operating within the
1.5 T magnetic field of a superconducting solenoid. 
Information from
a ring-imaging Cherenkov detector combined with energy-loss measurements in the 
SVT and DCH provide identification of charged kaon and pion candidates. 
The energies and locations of showers associated with photons are measured with a 
CsI(Tl) electromagnetic calorimeter.

\section{\boldmath Data Selection}

\begin{table*}[tbp]
\caption{List of reconstructed final states. Here $X$ indicates the rest of the event, with any number of charged 
or neutral particles.}
\label{tab:tab1}
\begin{center}
\vskip -0.2cm
\begin{tabular}{llll}
\hline \noalign{\vskip1pt}
Channel & Reaction  & \ \ \ $D^*$ decay mode    & \ \ \  $D$ decay mode \cr
\hline \noalign{\vskip2pt}
(1) & $e^+ e^- \to \Dz \Kp X $& & $\Dz \to \Km \pip $ \cr
(2) & $e^+ e^- \to \Dp \KS X $& & $\Dp \to \Km \pip \pip $\cr
(3) & $e^+ e^- \to \Dstarz \Kp X $& $\ \ \Dstarz \to \Dz \piz$ & $ \Dz \to \Km \pip$ \cr
(4) & $e^+ e^- \to \Dstarp \KS X $& $\ \ \Dstarp \to \Dp \piz$ & $ \Dp \to \Km \pip \pip $\cr
(5) & $e^+ e^- \to \Dstarp \KS X $& $\ \ \Dstarp \to \Dz \pip $& $\Dz \to \Km \pip$ \cr
(6) & $e^+ e^- \to \Dstarp \KS X $&$\ \ \Dstarp \to \Dz \pip$ &$ \Dz \to \Km \pip \piz$ \cr
(7) & $e^+ e^- \to \Dstarp \KS X $& $\ \ \Dstarp \to \Dz \pip$ & $\Dz \to \Km \pip \pip \pim $ \cr
\hline
\end{tabular}
\end{center}
\end{table*}

We reconstruct the inclusive processes~\cite{footnote} listed in
Table~\ref{tab:tab1}.
A particle identification algorithm is applied to all the tracks. Charged kaon identification has an average efficiency
of 90\% within the acceptance of the detector and
an average pion-to-kaon misidentification probability of 1.5\% per particle.

For all channels we perform a vertex fit for the $\Dz$ and $\Dp$ daughters
and require a $\chi^2$ probability greater than 0.1\%. For the $\pi^0$
candidates in channels (3),(4), and (6), we combine all photons with energy greater than 30 MeV in pairs,
perform a fit with 
a $\pi^0$ mass constraint, and require a $\chi^2$ probability greater than
1\%. For the $\Dz \to \Km \pip \piz$ decay channel we also perform 
a kinematic fit with a $\Dz$ mass constraint.  
We obtain $K^0_S \to \pip \pim$ candidates by means of a vertex fit and
require a $\chi^2$ probability greater than 2\%.  We accept only $K_S^0$ candidates with
decay length greater than 0.5 cm. To obtain $\Dstarp \KS$ candidates, where $\Dstarp \to \Dz \pip $, we combine fitted 
$\Dz$ and $\KS$ candidates with a \pip candidate using a vertex fit which constrains the overall vertex to be 
located in the interaction region, requiring a $\chi^2$ probability greater than 0.1\%. 
Similarly, for $\Dstarp \KS$ candidates where $\Dstarp \to \Dp \pi^0 $ we combine fitted 
$\Dp$, $\KS$ and $\piz$ candidates using a vertex fit which constrains the overall vertex to be 
located in the interaction region, requiring a $\chi^2$ probability greater than 0.1\%. 
Background from $e^+ e^- \to B \bar B$ events is removed by requiring the center of mass momentum $p^*$ of the $D K$ or $D^* K$ system to be greater 
than 3.3 \gevc.

To improve the signal to background ratio for channels with $D^0 \to \Km \pip$ we study the
distribution of the angle $\theta_{K^-}$ formed by the $K^-$ from $D^0$ decay
in the $K^- \pi^+$ rest frame with respect to the $K^- \pi^+$ direction in the
laboratory system. This distribution is expected to be flat.
We observe an accumulation of combinatorial background close to $\cos \theta_{K^-}=1$. We improve the signal to background
ratio by requiring $\cos \theta_{K^-}<0.9$.

\begin{figure}
\vskip -0.15in
\includegraphics[width=\columnwidth]{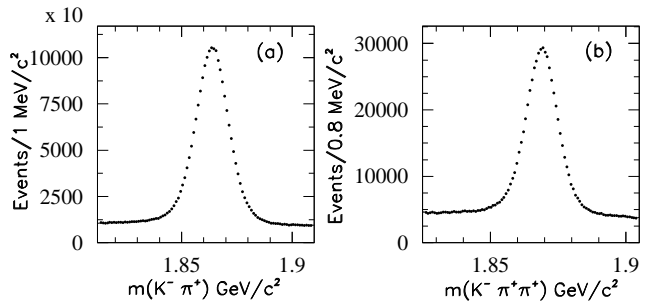}
\caption{\label{fig:fig1}
(a) $\Km\pip$, (b) $\Km\pip\pip$  
mass distributions for all candidate events in channels (1) and (2) 
respectively.}
\end{figure}

To improve the signal to background ratio for $\Dp \to \Km \pip \pip$, we compare the $\Dp$ three-momentum and its flight
direction and define $d_{xy}$ as the signed projected distance in the transverse plane.
Background events are removed by requiring $d_{xy}>0$.
The resulting $\Km \pip$ and $\Km \pip \pip$ invariant mass spectra for candidates in channels (1) 
(where we require in addition a reconstructed \Kp) and (2) (where we require in addition a reconstructed $\KS$) are shown in Fig.~\ref{fig:fig1}. There are on average 1.01 candidates per selected event in
both samples, and all candidates are retained for further analysis.

We fit the $\Km \pip$ and $\Km \pip \pip$ invariant mass spectra using a linear background and a single Gaussian peak obtaining $\sigma_{\Dz} = 7.6$ $\mevcc$ and $\sigma_{\Dp}= 6$ $\mevcc$.
The signal region is defined within 
$\pm 2 \sigma$ while sideband regions are defined within $(-6\sigma, -4\sigma)$ and $(4\sigma, 6\sigma)$.
The $\Dz$ signal region contains 1.98 $\times 10^6$ combinations with a purity
$P=N_S/(N_S+N_B)=0.84$, where $N_S$ ($N_B$) is the number of signal
(background) combinations. 
The $\Dp$ signal region contains 0.58 $\times 10^6$ combinations with a purity $P=0.75$. 
 
\section{\boldmath Study of the $D K$ systems}

We first study the $\Dz \Kp$ and $\Dp \KS$ mass spectra.
In an inclusive environment the $\Dz$ and  $\Dp$ can come from $D^*$ decays.
Candidate $\Dz \Kp$ pairs where the $\Dz$ is a $D^*$-decay product are identified by 
forming $\Dz\pip$, $\Dz\piz$, and $D\gamma$ combinations and requiring that the invariant-mass difference 
between one of those combinations and the $\Dz$ be within $\pm 2\sigma$ of the known $D^*-D$ 
mass difference.
Events belonging to these
possible reflections (except for $D^{*0} \to D^0 \gamma$ events, which could not be 
isolated cleanly) are removed.
In the same way, $\Dp \KS$ combinations where the $\Dp \piz$ and $\Dp$ invariant-mass difference 
is found to be within $\pm 2\sigma$ of the known $D^*-D$ 
mass difference are removed.

We also study the distribution of $\theta_{\Kp}$ ($\theta_{\KS}$), the angle between the $\Kp$ ($\KS$) direction in the 
$D K$ rest frame and the $D K$ direction in the laboratory frame. We expect the distribution of this 
angle to be symmetric around zero~\cite{cleo}, but we observe an accumulation of
combinatorial background close to $\cos \theta_{K}= -1$. 
Due to the jetlike nature of the reaction $e^+ e^- \to c \bar c$ we interpret 
this background as due to combinations for which the $K$ comes from the jet opposite to the $D$ meson. We therefore
apply a conservative cut requiring  $\cos \theta_{K}>-0.8$.  

\begin{figure*}
\vskip -0.15in
\includegraphics[width=16cm]{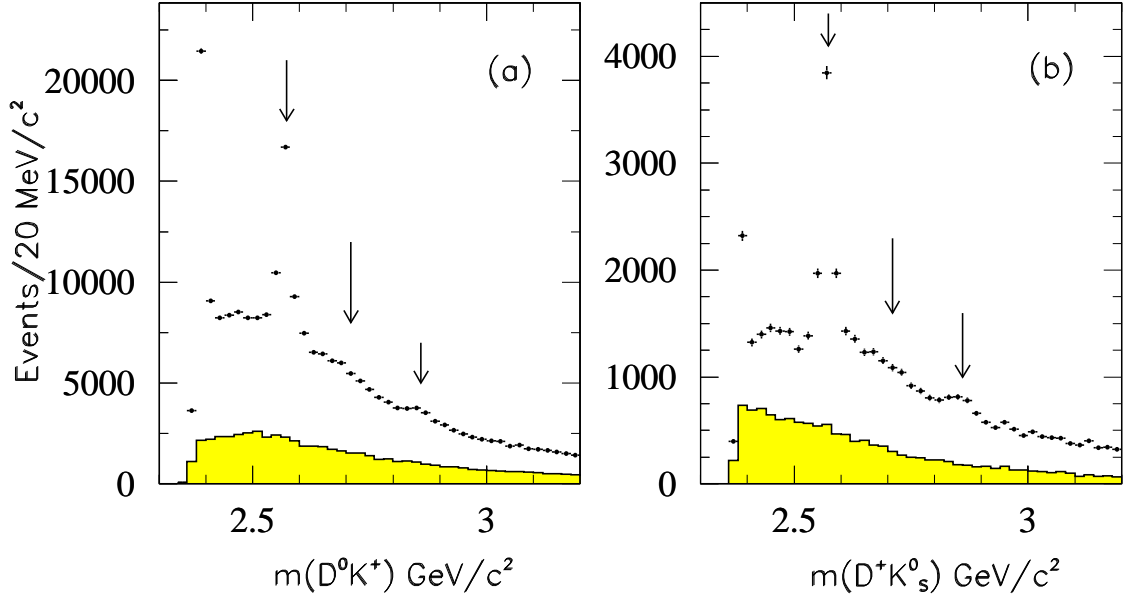}
\vskip -0.15in
\caption{\label{fig:fig2} 
$D K$ invariant mass distributions for (a) $D^0_{K^-\pi^+} K^+$,
and (b) $D^+_{K^-\pi^+\pi^+} K^0_s$.
Shaded histograms represent the $D$ mass sideband regions. The arrows indicate the expected positions of the $D_{s2}^*(2573)^+$, $\Dsa$, and
$\Dsb$.
}
\end{figure*}

The resulting $\Dz \Kp$ and $\Dp \KS$ mass spectra are shown in Fig.~\ref{fig:fig2}.
To improve the mass
resolution, the nominal $D$ mass and the reconstructed 3-momentum
are used to calculate the $D$ energy for channels (1) and (2).
The two mass spectra in Fig.~\ref{fig:fig2} present similar features.
The single bin peak at 2.4 \gevcc results from  
decays of $D_{s1}(2536)^+$  
to $D^{*0} K^+$ or $D^{*+} K^0_S$ in which the $\pi^0$ or $\gamma$ from the 
$D^*$ decay is missed. Since the $D_{s1}(2536)^+$ is believed to have $J^P = 1^+$,
         decay to $D K$ is forbidden by angular momentum and parity
         conservation. 
We also observe a prominent narrow signal due to the $D_{s2}^*(2573)^+$, 
a broad structure centered at the mass of the $\Dsa$, and a narrower structure at the 
position of the $\Dsb$.

\begin{figure*}
\begin{center}
\vskip -0.15in
\includegraphics[width=12cm]{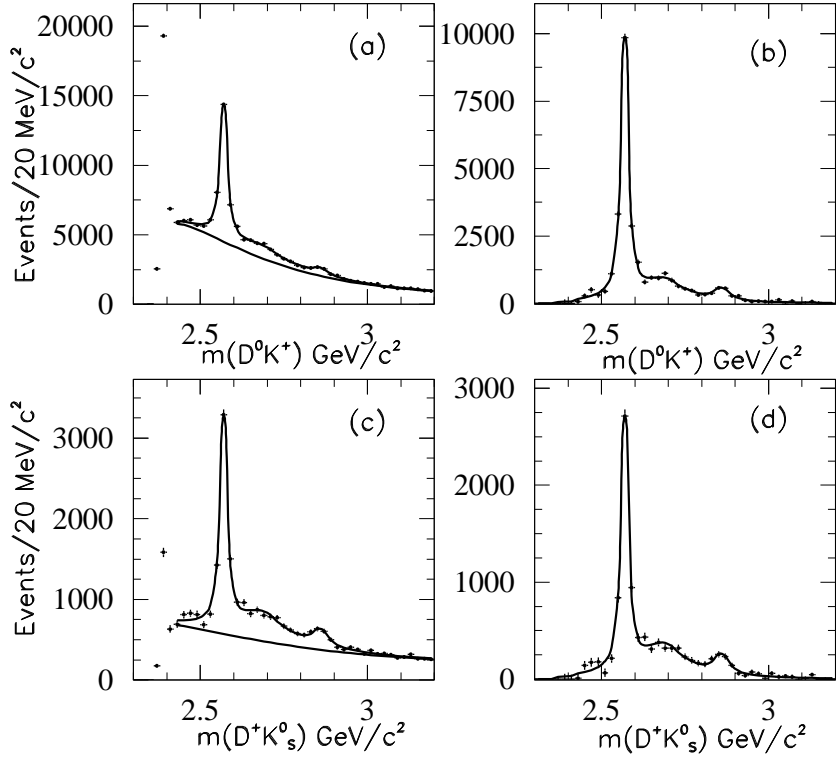}
\end{center}
\vskip -0.15in
\caption{\label{fig:fig3} Sideband-subtracted $D K$ invariant mass 
distributions for 
(a) $D^0_{K^-\pi^+} K^+$, (c) $D^+_{K^-\pi^+\pi^+}K^0_s$; (b) and (d) show the fitted-background-subtracted
mass spectra.
The curves show the functions described in the text.
}
\end{figure*}

We perform a simultaneous binned $\chi^2$ fit to the two sideband-subtracted $D K$ mass spectra shown in Fig.~\ref{fig:fig3}(a) and Fig.~\ref{fig:fig3}(c).
The fit range extends from 2.42 \gevcc to 3.2 \gevcc (the lower bound is chosen to exclude the
$D_{s1}(2536)^+$ reflection). The background for each of the two $D K$ mass
distributions is described by a threshold function: 
$(m - m_{\rm th})^{\alpha}~e^{-\beta m-\gamma m^2-\delta m^3}$ 
where $m_{\rm th}=m_{D}+m_K$. 
In this fit, the $D_{s2}^*(2573)^+$, $\Dsa$ and $\Dsb$ peaks are
described with
relativistic Breit-Wigner lineshapes where spin-2 is assumed for $D_{s2}^*(2573)^+$, spin-1 
for $\Dsa$ and spin-0
for $\Dsb$. The Breit-Wigner function includes Blatt-Weisskopf form 
factors~\cite{dump}. 
Mass resolution estimates are obtained using simulated events. We obtain single-Gaussian $\sigma$'s of 2.7 \mevcc and 3.6 \mevcc 
at $D K$ masses
of 2.71 and 2.86 \gevcc respectively. Since the width values for the resonances present in the $D K$ mass spectra are much larger than 
these, resolution effects are ignored.

The result of the fit is shown in Figs.~\ref{fig:fig3}(a) and ~\ref{fig:fig3}(c), and the parameter values obtained 
are summarized in Table~\ref{tab:tab2} (Fit A).
Figures ~\ref{fig:fig3}(b) and \ref{fig:fig3}(d) show also the fitted-background-subtracted $D^0_{K^-\pi^+} K^+$, and $D^+_{K^-\pi^+\pi^+} K^0_s$
invariant mass distributions.  
The fitted parameter values for the $\Dsb$ state are in agreement with our previous measurement~\cite{babar_ds},
while the central value of the $\Dsa$ mass is slightly higher than before. 

\begin{table*}[tbp]
\caption{The $\chi^2$/NDF and resonance parameter values obtained from the fits to the $D K$ and
  $D^*K$ mass spectra. Masses and widths are given in units of \mevcc and \mev, respectively. Uncertainties are statistical only.
}
\label{tab:tab2}
\begin{center}
\vskip -0.2cm
\begin{tabular}{l@{\hspace{3mm}}c@{\hspace{3mm}}c@{\hspace{3mm}}c@{\hspace{3mm}}c}
\hline \noalign{\vskip2pt}
Fit & $\chi^2/{\rm NDF}$ & $\Dsa$  & $\Dsb$ & $\Dsc$ \cr
\hline \noalign{\vskip2pt}
A ($D K$) & 85/56 &  $m =2710.0 \pm 3.3$   & $m=2860.0 \pm 2.3$ & \cr
 &        & $\Gamma= 178 \pm 19$ &  $\Gamma=  53 \pm 6$ & \cr
\hline \noalign{\vskip2pt}
B ($D^*K$)& 51/33 & $m = 2712 \pm 3$ & $m=2865.2 \pm 3.5 $  & $m=3042 \pm 9$ \cr
 & & $\Gamma =103 \pm 8 $ & $\Gamma=44 \pm 8.3$ & $\Gamma= 214 \pm 34 $ \cr
C ($D K+D^*K$) & 147/91 & $m = 2710 \pm 3$ & $m=2860 \pm 2 $  & $m=3045 \pm 8$ \cr
 & & $\Gamma =152 \pm 7 $ & $\Gamma=52 \pm 5$ & $\Gamma= 246 \pm 31 $ \cr
 & & & $m=2866 \pm  3$ & \cr
 & & & $\Gamma=43 \pm  6$ & \cr
D ($D K+D^*K$) & 149/93 & $m = 2710 \pm 2$ & $m=2862 \pm 2 $  & $m=3044 \pm 8$ \cr
 & & $\Gamma =152 \pm 7 $ & $\Gamma=48 \pm 3$ & $\Gamma= 239 \pm 35 $ \cr
\hline \noalign{\vskip2pt}
E ($D^*K$) & 65/38 & $m = 2716.7 \pm 2.5$ & & \cr
 & & $\Gamma = 108 \pm 5 $ & & \cr
F ($D^*K$) & 39/34 & & & $m = 3047 \pm 12$ \cr
 & & & & $\Gamma = 216 \pm 46 $  \cr
\hline
\end{tabular}
\end{center}
\end{table*}

\section{\boldmath Study of the $D^* K$ system}

\begin{figure*}
\vskip -0.15in
\includegraphics[width=14cm]{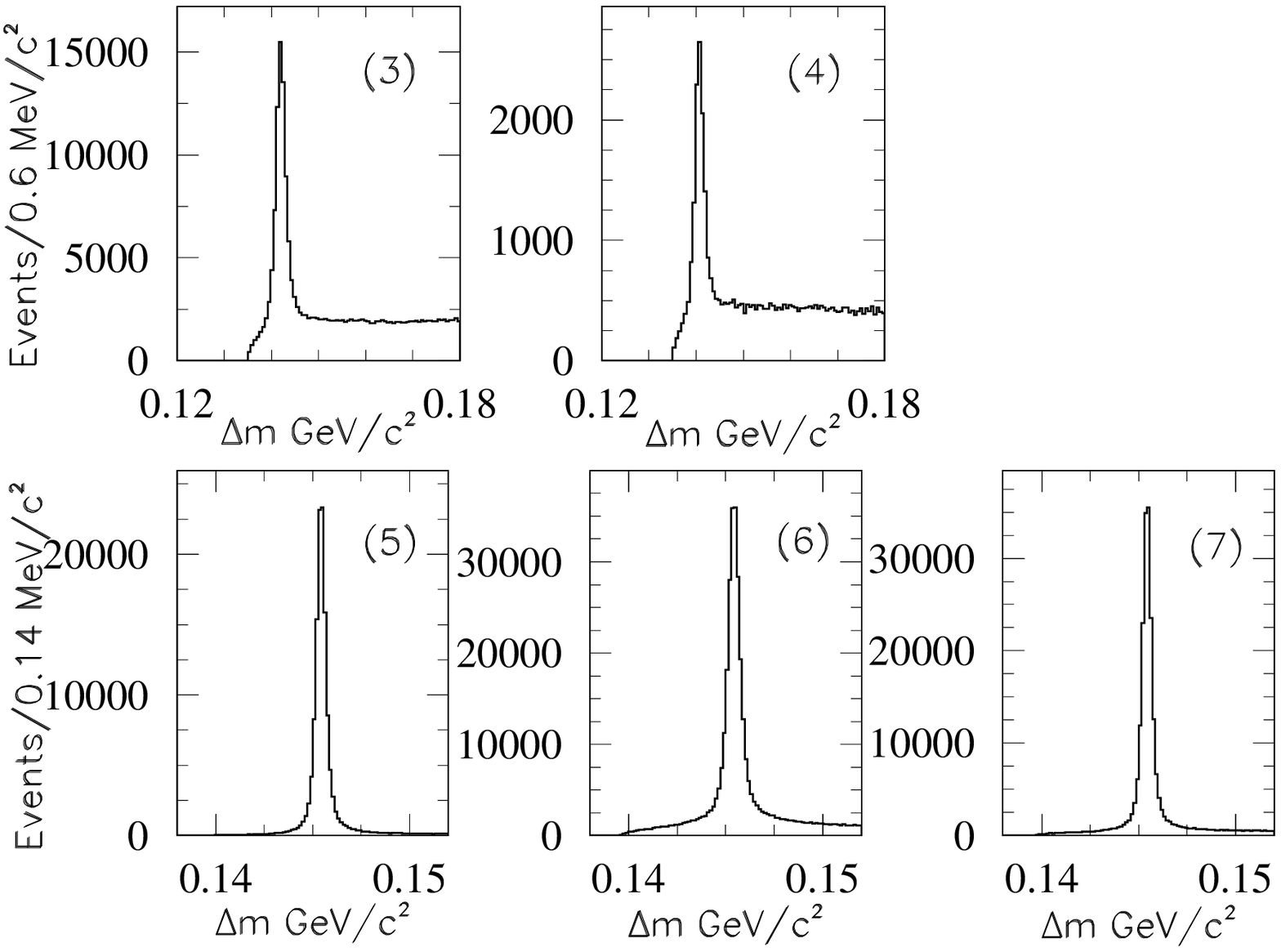}
\vskip -0.15in
\caption{\label{fig:fig4} 
The $\Delta m$ distributions for channels (3)-(7) after applying the corresponding $D$-candidate mass selection criteria. 
}
\end{figure*}

The $\Delta m = m(D \pi) - m(D)$ distributions for the five channels (3)-(7), 
are shown
in Fig.~\ref{fig:fig4}. Backgrounds are small for channels (5)-(7) but larger
for channels (3)-(4). 
Table~\ref{tab:tab3} gives the fitted parameter values of the $\Delta m$ distributions
together with purities and the definitions of signal and sideband regions. The values of $\sigma$ reported
in Table~\ref{tab:tab3} are obtained from fits performed using a polynomial 
background and a single Gaussian. 

\begin{table*}[tbp]
\caption{Fitted parameters of the $\Delta m$ distributions together with purities and definitions of the regions used for signal 
and background.}
\label{tab:tab3}
\begin{center}
\vskip -0.2cm
\begin{tabular}{lcccccc}
\hline \noalign{\vskip1pt}
Channel & mass  & $\sigma$  & purity (\%) & Signal region & Sideband region \cr
 & $\mevcc$ & $\mevcc$ & & & \cr
\hline \noalign{\vskip1pt}
(3) $\Delta m(\Dz_{K^-\pi^+} \piz) $ &  142.02 & 1.08 & 83.3 & $\pm 2.5 \sigma$ & $10 \sigma-15 \sigma$ \cr
\noalign{\vskip1pt}
(4) $\Delta m(\Dp_{K^-\pi^+\pi^+} \piz)$ & 140.63 & 0.893 & 76.6 & $\pm 2.5 \sigma$ & $10 \sigma-15 \sigma$ \cr
\noalign{\vskip1pt}
(5) $\Delta m(\Dz_{K^-\pi^+} \pip)$ &145.43 & 0.288 & 94.9 & $\pm 5 \sigma$ & $12 \sigma-22 \sigma$ \cr
\noalign{\vskip1pt}
(6) $\Delta m(\Dz_{K^-\pi^+\pi^0} \pip)$ &145.43 & 0.351 & 87.1 & $\pm 3 \sigma$ & $10 \sigma-16 \sigma$ \cr
\noalign{\vskip1pt}
(7) $\Delta m(\Dz_{K^-\pi^+\pi^+\pi^-} \pip)$ &145.43 & 0.266 & 90.5 & $\pm 5 \sigma$ & $12 \sigma-22 \sigma$ \cr
\noalign{\vskip1pt}
\hline
\end{tabular}
\end{center}
\end{table*}

We have also studied the distributions of the angle $\theta^*_{\Kp}$ ($\theta^*_{\KS}$) between the $\Kp$ ($\KS$) direction in the 
$D^*K$ rest frame and the $D^*K$ direction in the laboratory frame. We expect the distributions of this 
angle to be symmetric around zero for signal but we observe an accumulation of combinatorial background close to $\cos \theta^*_{K}=-1$. 
As in the case of the $D K$ system, we interpret 
this as being due to combinations for which the $K$ comes from the jet opposite to that yielding the $D^*$ meson. We therefore
apply the conservative selection criterion $\cos \theta^*_{K}>-0.8$.  

\begin{figure*}
\vskip -0.15in
\includegraphics[width=14cm]{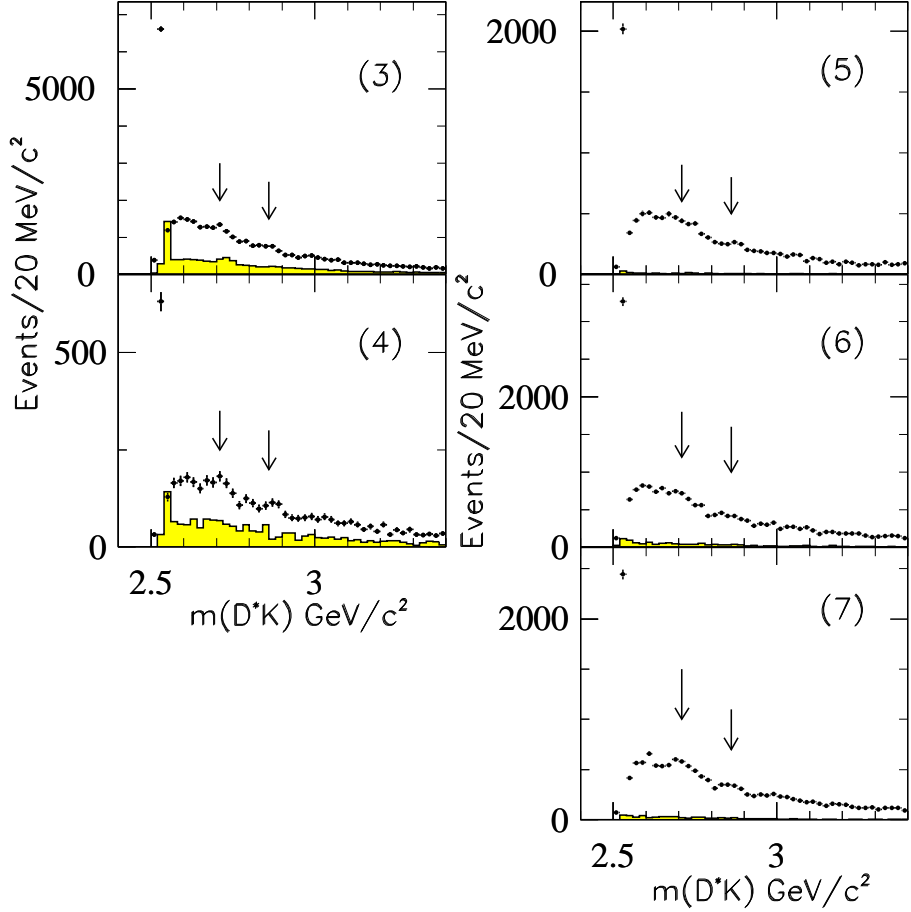}
\vskip -0.15in
\caption{\label{fig:fig5} 
The $m(D^*K)$ distributions for channels (3)-(7); the shaded histograms show the mass spectra from the 
$D^*$ $\Delta m$ sideband regions.
The arrows indicate the expected positions of the $\Dsa$ and $\Dsb$.}
\end{figure*}

Figure~\ref{fig:fig5} shows the reconstructed $D^*K$ mass spectra for channels (3)-(7) of Table~\ref{tab:tab1}. 
The shaded distributions represent
the background estimates from the $\Delta m$
sideband regions. Each
distribution shows a narrow spike at threshold due to the $D_{s1}(2536)^+$ meson. We also observe structures around 2.71 and 
2.86 \gevcc. 

We have compared these mass spectra with those from generic $e^+ e^- \to \bar c c$ Monte Carlo events.
These events were
generated using a detailed 
detector simulation and subjected to the same reconstruction
and event-selection procedure as used for the data.
We find that the simulation underestimates the $D_{s1}(2536)^+$ and
$D_{s2}^*(2573)^+$ signals relative to the background. No such discrepancy 
is found in the study of non-strange final states, therefore we attribute 
this effect
to poor knowledge of the strange-charmed meson cross sections~\cite{lafferty}. We apply weights 
to the $D_{s1}(2536)^+$ and $D_{s2}^*(2573)^+$ production in the Monte Carlo in order to obtain better agreement 
with the data. 

Figure~\ref{fig:fig5}(3) shows the presence of a peaking background in channel (3) around 2.7 \gevcc.
Using the Monte Carlo data we identify this reflection, which is present in the signal and
the sideband regions, 
as being due to  the $D_{s2}^*(2573)^+$.  Combinations of $D K$
originating from this narrow peak associate with a random $\pi^0$ to produce a relatively narrow
structure in the 2.7 \gevcc region.  Our Monte Carlo study verifies that this reflection is almost completely 
removed by performing the $\Delta m$ sideband subtraction.

\begin{figure*}
\vskip -0.15in
\includegraphics[width=14cm]{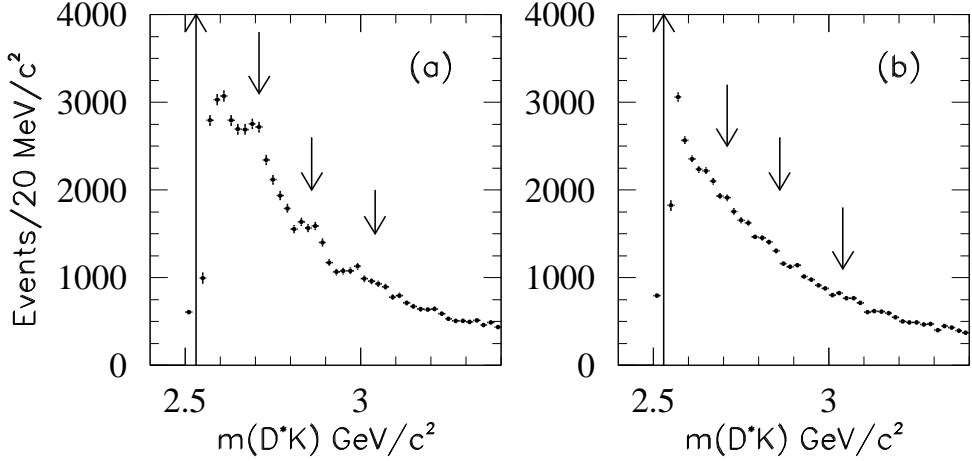}
\vskip -0.15in
\caption{\label{fig:fig6} 
The sideband-subtracted $m(D^*K)$ distributions summed over channels (3)-(7). (a) Data, (b) Monte Carlo. The arrow
near threshold indicates the position
of the peak due to the $D_{s1}(2536)^+$, which is off-scale. The other arrows indicate the expected positions of the $\Dsa$, $\Dsb$, and $\Dsc$.
}
\end{figure*}

The total $D^*K$ mass spectrum, $\Delta m$-sideband-subtracted and summed over channels (3)-(7), is shown in Fig.~\ref{fig:fig6} 
and compared with that obtained from
Monte Carlo simulations. 
The $D^*K$ mass spectrum, above the $D_{s1}(2536)^+$ signal, shows the presence of structures around 2.71, 2.86, and 3.04 \gevcc. 
Corresponding resonance contributions are
not included in the Monte Carlo simulation. Since such enhancements 
are not evident in the Monte Carlo $D^*K$ mass spectrum, we conclude that these structures are not produced by reflections
from known resonances. Monte Carlo simulations also show that these enhancements are not due to reflections from the
$D_{sJ}$ resonances observed in the $DK$ mass spectrum.

A structure close to 2.57 \gevcc is seen in the Monte Carlo and is due to the decay $D_{s2}^*(2573)^+ \to D^*K$, included in the simulation.
However the data do not show evidence for such a decay.

\section{\boldmath Fits to the $D^*K$ mass spectrum}

We perform a binned minimum $\chi^2$ fit to the combined $D^*K$ mass spectrum. The fit is performed in the region (2.58-3.48) \gevcc.
The background is parametrized as   
$e^{-\beta m-\gamma m^2-\delta m^3}$,
which provides a good description of the Monte Carlo in the same mass range.
The $D_{sJ}^+$ peaks are described with relativistic Breit-Wigner lineshapes. Here and in the following fits we assume 
$J^P=1^-$ and $J^P=3^-$ respectively for \Dsa and \Dsb decays to $D K$. For the $D^*K$ system, we use an angular momentum 
$L=1$, $L=3$, and $L=0$ for \Dsa, \Dsb, and \Dsc respectively.
Average mass resolutions are 2.5 \mevcc and 3.5 \mevcc  at $D^*K$ masses
of 2.71 and 2.86 \gevcc respectively. Since the width values for the resonances present in the $D^*K$ mass spectra are much larger than 
these, resolution effects are ignored.

\begin{figure*}
\vskip -0.15in
\includegraphics[width=14cm]{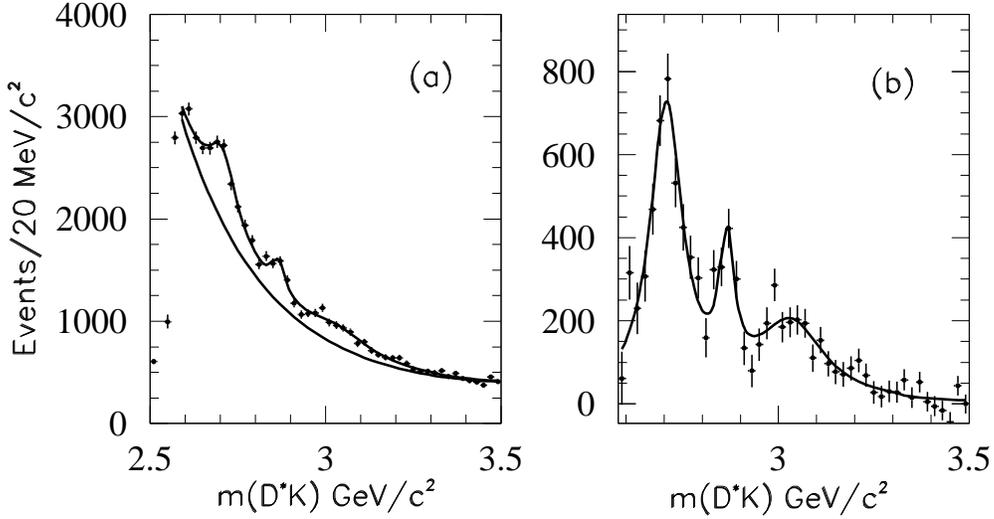}
\vskip -0.15in
\caption{\label{fig:fig7} (a) Fit to the $D^* K$ invariant mass spectrum. (b) Residuals after subtraction of the fitted background.
The curves are described in the text.
}
\end{figure*}

We observe the presence, above the $\Dsa$ and $\Dsb$, of a new broad structure peaking at 3.04 $\gevcc$.
The resonance parameters resulting from the fit are given in Table~\ref{tab:tab2} (Fit B) and the corresponding fitted curves are shown
in Fig.~\ref{fig:fig7}.
Modifying the background to include an extra term in the exponential does not improve the fit significantly. 
We find that the width of the $\Dsa$ differs somewhat between the $D K$ and $D^*K$ fits, while the parameter values for the structure 
at 2.86 $\gevcc$ in the $D^*K$ mass spectrum are consistent with those of the $\Dsb$ obtained from the $D K$ mass spectrum. 

We next repeat the fits removing the resonances one by one from the PDF. 
We compute the statistical significance of each structure as $\sqrt{\Delta \chi^2}$ where $\Delta \chi^2$ is the 
difference in the fit $\chi^2$ with and without the resonance included, taking into account the variation in the number of parameters
($\Delta{\rm NDF}=3$). 
We obtain statistical significances of 12.4, 6.4, and 6.0 standard deviations for the $\Dsa$, $\Dsb$, and 
$\Dsc$ respectively.

We then perform simultaneous fits to the two $D K$ mass spectra and to the total $D^* K$ mass spectrum to better constrain the $\Dsa$ parameters.
We first test the possibility that the structure around 2.86 \gevcc in the $D^*K$ mass spectrum is different from the $\Dsb$ state observed in the $D K$ mass spectrum by adding to the fit two new parameters. The results from the 
fit are summarized in Table~\ref{tab:tab2} (Fit C). 
We find the parameters of the $\Dsb$ in the $D K$ mass spectrum consistent with those measured in the $D^*K$ mass spectrum.

Assuming therefore that we are observing the same state, we constrain the $\Dsb$ parameters to be the same in both the $D K$ and $D^*K$ mass spectra
(Fit D).  Taking this as our reference fit, we obtain the following parameters
for the three states:

\begin{equation}
\begin{split}
m(\Dsa) = 2710 \pm 2_{\sta}(^{+12}_{-7})_{\sys}\ \mevcc \\
\Gamma=149 \pm 7_{\sta}(^{+39}_{-52})_{\sys} \ \mev,
\end{split}
\end{equation}

\begin{equation}
\begin{split}
m(\Dsb) = 2862 \pm 2_{\sta}(^{+5}_{-2})_{\sys}\ \mevcc \\
\Gamma=48 \pm 3_{\sta}\pm 6_{\sys} \ \mev,
\end{split}
\end{equation}

\begin{equation}
\begin{split}
m(D_{sJ}(3040)) = 3044 \pm 8_{\sta}(^{+30}_{-5})_{\sys}\ \mevcc \\
\Gamma=239 \pm 35_{\sta}(^{+46}_{-42})_{\sys} \ \mev.
\end{split}
\end{equation}

Here systematic uncertainties take into account the range of values obtained in different fits,
including fits to the spectra obtained after modifying the $p^*$ selection criterion, modifying the $\Delta m$ criteria, and
removing the $\cos \theta_K$ requirement. They also account for uncertainties in the spin assignment.

\section{\boldmath Angular analysis}

\begin{figure*}
\vskip -0.15in
\includegraphics[width=14cm]{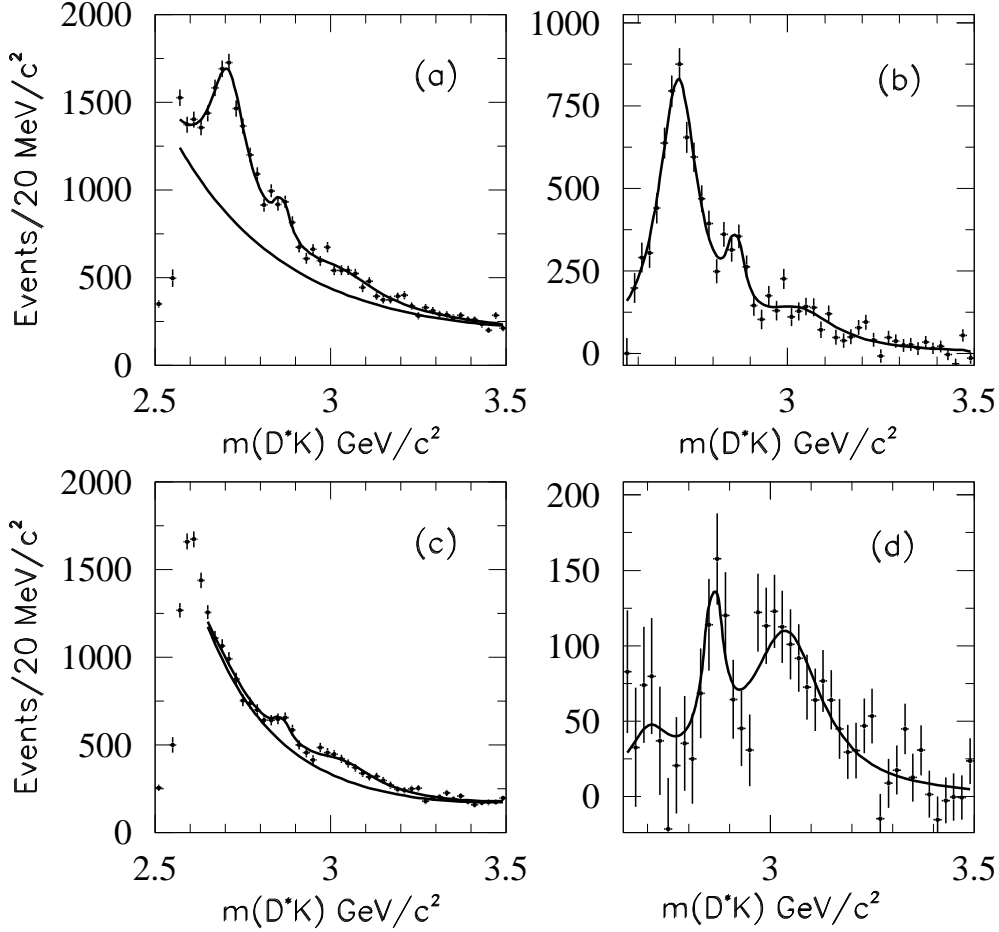}
\vskip -0.15in
\caption{\label{fig:fig8} Fits to the $D^* K$ invariant mass spectra for (a) $\left| \cos \theta_h \right| < $ 0.4; 
(c) $\left| \cos \theta_h \right| >$ 0.4. (b) and (d) show the data after the fitted background is subtracted.}
\end{figure*}

Since we observe both $\Dsa$ and $\Dsb$ decays to both $D K$ and $D^*K$, we
assume they have natural parity $J^P=1^-,2^+,3^-,...$ ($J^P=0^+$ is ruled out
because of the $D^*K$ decay).  We further test this hypothesis using angular analysis.
We compute the helicity angle $\theta_h$
as the angle formed by the $\pi$ from the $D^*$ decay with respect to the
kaon, in the $D^*$ rest frame. The angular distribution for natural parity is expected to be~\cite{avery}
\begin{equation}
\frac{d N}{d \cos \theta_h} \cong  1 - \cos^2 \theta_h \ ,
\end{equation}
since, for the parity and angular momentum
                              conserving decay of such a parent state,
                              the coupling in the parent rest frame to
                              the helicity-0 $D^*$ state involves a
                              vanishing Clebsch-Gordan coefficient.
Figure~\ref{fig:fig8} shows the $D^*K$ mass spectrum separated for $\left| \cos \theta_h \right| < 0.4$ ((a),(b)) and $\left| \cos \theta_h \right| > 0.4$ ((c),(d)). We clearly 
observe an enhanced signal to background rate for the  $\Dsa$ in the $\left| \cos \theta_h \right| < 0.4$ region. 
This does not hold for the $\Dsc$. 
The non-observation of $\Dsc \to D K$ also suggests unnatural parity for this state. 

The mass spectra separated according to the value of $\left| \cos \theta_h \right|$ allow a better determination of the $\Dsa$ and $\Dsc$ parameters.
When fitting the 
$\left| \cos \theta_h \right| < 0.4$ data (Fit E) 
we fix the $\Dsb$ and $\Dsc$ shape parameters to those resulting from the
simultaneous fit of the $D K$ and $D^*K$ mass spectra (Fit D). 
In fitting the  $\left| \cos \theta_h \right| > 0.4$ data (Fit F),
we fix the parameters of $\Dsa$ and $\Dsb$ to those from Fit D. 
The resulting $\Dsa$ and $\Dsc$ parameters are given in Table~\ref{tab:tab2} and the fit results are shown by the curves in Fig.~\ref{fig:fig8}.

We have studied the $\cos \theta_h$ distributions for the $\Dsa$ and $\Dsb$ by producing $D^* K$ mass spectra in six intervals of $\cos \theta_h$. The
mass spectrum in each interval was fitted with fixed resonance parameters. However these values have all been varied within their statistical and 
systematic errors. The efficiencies as a function of $\cos \theta_h$ in the two mass regions are obtained from Monte Carlo simulation
of the five channels involved in the analysis. We find that the efficiency is almost uniform as a function of $\cos \theta_h$ with a small
slope which we parametrize by a linear function.

\begin{figure}
\vskip -0.15in
\includegraphics[width=10cm]{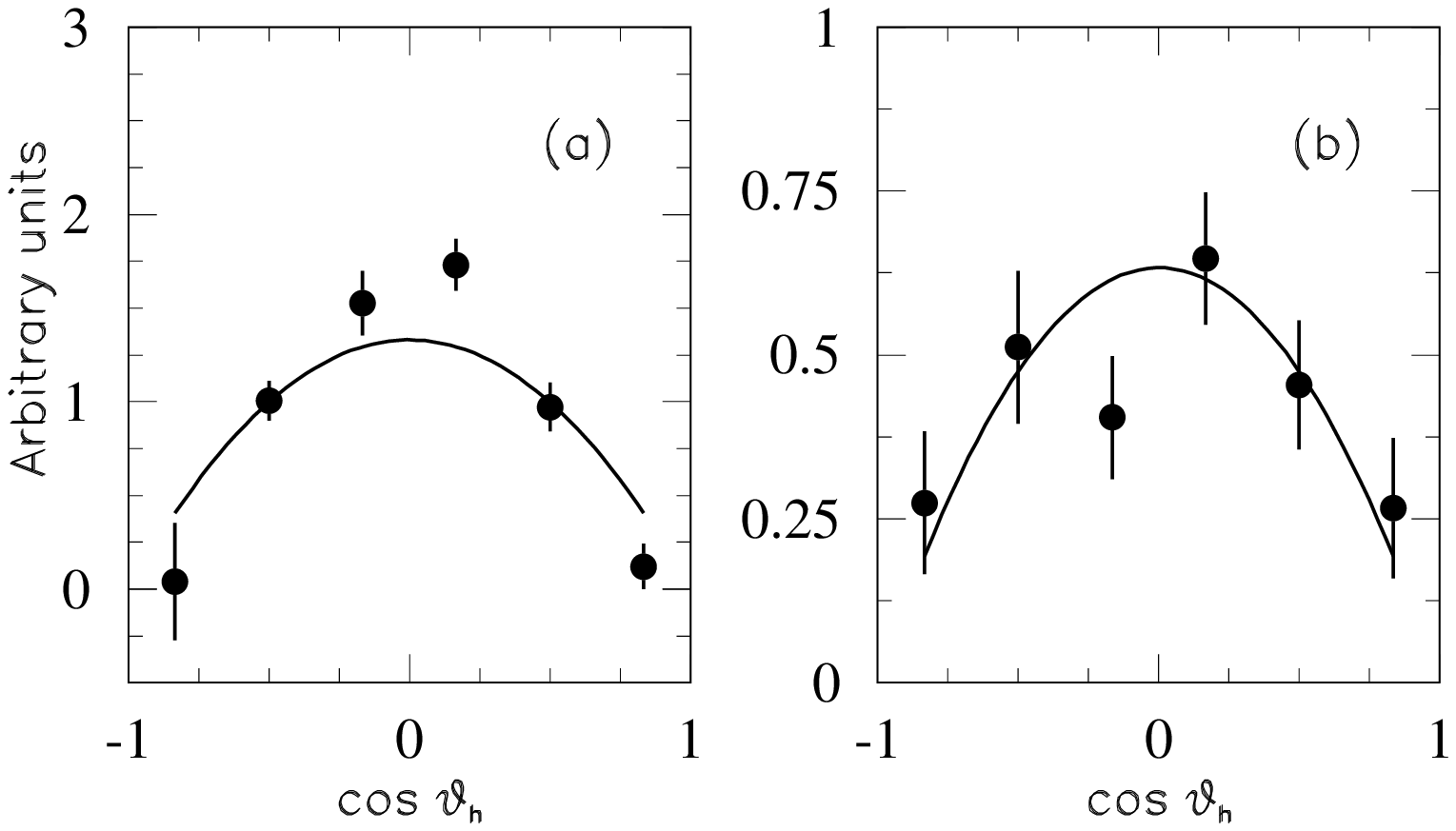}
\vskip -0.15in
\caption{\label{fig:fig9} Distributions in $cos \theta_h$ for (a) the $\Dsa$ and (b) the $\Dsb$.}
\end{figure}

The efficiency corrected $\Dsa$ and $\Dsb$ yields are plotted in Fig.~\ref{fig:fig9}, together with the normalized expectations for natural 
parity. The curves have $\chi^2/{\rm NDF}$ of 18.7/5 and 6.3/5 respectively. The large $\chi^2$ obtained for the $\Dsa$ is related to the large
uncertainties in the background parametrization. Other spin hypotheses have been tested but they give much larger $\chi^2$ values.
We conclude that both states are consistent with having natural parity.
We do not perform a similar analysis for $\Dsc$ because of the large
uncertainties arising from fitting a very broad resonance with limited statistics. 

\section{\boldmath Branching fractions}

From Table~\ref{tab:tab1} it can be seen that it is possible to obtain ratios of branching fractions with reduced systematic uncertainties, for $\Dsa$ and $\Dsb$ by using
channels (3),(1) and (4),(2) respectively.

These ratios are computed as:

\begin{equation}
r_i = \frac{N(D^*_{sJ}\to D^*K)}{N(D^*_{sJ}\to D K)}\frac{\epsilon(D^*_{sJ}\to D K)}{\epsilon(D^*_{sJ}\to D^*K)}
\end{equation}

\noindent
where the $N(D^*_{sJ})$ are the signal yields and the $\epsilon(D^*_{sJ})$ are the corresponding efficiencies, and $i$=1,4.
We note that the only difference between numerator and
denominator final states is the presence of an extra $\pi^0$ from the $D^*$ decay.

Assuming a constant total width,
the yields are obtained by fitting the $D K$ and $D^*K$ mass spectra using the same $\Dsa$ and $\Dsb$ parameters,
and are summarized in
Table~\ref{tab:tab4}.  Efficiencies are evaluated using Monte Carlo simulations, and only 
the ratio of efficiencies

\begin{equation}
\epsilon_r = \frac{\epsilon(D^*_{sJ}\to D^*K)}{\epsilon(D^*_{sJ}\to D K)}
\end{equation}

\noindent
is involved in the measurement. This ratio is consistent with being uniform as a function of  $D K(D^*K)$ mass and 
its values are reported in Table~\ref{tab:tab4}.

\begin{table*}
\caption{Information related to the evaluation of the ratio of branching fractions for the $D^*_{sJ}$ resonances.}
\label{tab:tab4}
\begin{center}
\vskip -0.2cm
\begin{tabular}{lcccc}
\hline \noalign{\vskip1pt}
Decay & N & $\epsilon_r$ & $D^*$ B.F.(\%) & $r_i$  \cr
\hline \noalign{\vskip1pt}
$\Dsa \to D^0 K^+$  & 6469 $\pm$  425 & & & \cr 
$\Dsa \to D^{*0} K^+$ & 1247 $\pm$  173 & 0.353 $\pm$  0.005  & 61.9 $\pm$ 2.9 & 0.88 $\pm$  0.14 \cr
$\Dsb \to D^0 K^+$  & 1826 $\pm$ 158 & & & \cr
$\Dsb \to D^{*0} K^+$ & 415 $\pm$ 55 & & & 1.04 $\pm$ 0.17 \cr
\hline \noalign{\vskip1pt}
$\Dsa \to D^+ K^0_S$  & 2442 $\pm$  179 & & & \cr
$\Dsa \to D^{*+} K^0_S$ & 258 $\pm$  75 & 0.301 $\pm$ 0.009 & 30.7 $\pm$ 0.5 & 1.14 $\pm$ 0.39 \cr
$\Dsb \to D^+ K^0_S$  & 781 $\pm$  83 & & & \cr
$\Dsb \to D^{*+} K^0_S$ & 100 $\pm$  23 & & & 1.38 $\pm$ 0.35 \cr
\hline
\end{tabular}
\end{center}
\end{table*}

Systematic uncertainties are summarized in Table~\ref{tab:tab5}. The $D^*_{sJ}$ parameters have been varied within their
statistical and systematic errors, the $p^*$ cut has been changed to 3.1 and 3.5 \gevc, the $\cos \theta_K$ cuts have been 
increased to $-0.6$. The systematic error arising from Monte Carlo statistics
appears as the error on the ratio between 
the efficiencies. The error on the $D^*$ branching fractions is obtained from Ref.~\cite{pdg}. The shape of the 
background has been modified by adding an extra term in the exponential; its contribution to the total 
error is found to be negligible. Finally the $D^*$ $\Delta m$ signal region has been reduced to $\pm 2 \sigma$.

\begin{table}[tbp]
\caption{Systematic uncertainties in the evaluation of the ratio of branching fractions.}
\label{tab:tab5}
\begin{center}
\vskip -0.2cm
\begin{tabular}{lccccccc}
\hline \noalign{\vskip1pt}
Ratio & $D_{sJ}$ & MC  & $D^*$  & $p^*$ & $\cos \theta_K$  & $\Delta m$ & Total \cr
 & parameters & statistics & B.F. & cut  &cut  & cut \cr
\hline \noalign{\vskip1pt}
$r_1$ & 0.030 & 0.013 & 0.042 & 0.018 & 0.120 & 0.044 & 0.14 \cr
$r_2$ & 0.050 & 0.015 & 0.050 & 0.066 & 0.175 & 0.004 & 0.20 \cr
$r_3$ & 0.077 & 0.033 & 0.018 & 0.042 & 0.128 & 0.160 & 0.23 \cr
$r_4$ & 0.106 & 0.040 & 0.022 & 0.009 & 0.328 & 0.345 & 0.49 \cr
\hline
\end{tabular}
\end{center}
\end{table}

We obtain the following ratios of branching fractions:

\begin{equation}
\begin{split}
r_1 = \frac{
  {\cal B}(\Dsa \to D^{*0}K^+)}{{\cal B}(\Dsa \to D^0K^+)
} = \\
  0.88 \pm  0.14_{\sta} \pm  0.14_{\sys}
\end{split}
\end{equation}

\begin{equation}
\begin{split}
r_2 = \frac{
  {\cal B}(\Dsb \to D^{*0}K^+)}{{\cal B}(\Dsb \to D^0K^+)
} = \\
 1.04 \pm 0.17_{\sta} \pm 0.20_{\sys}
\end{split}
\end{equation}

where $D^{*0} \to D^0 \pi^0$, and:

\begin{equation}
\begin{split}
r_3 = \frac{
  {\cal B}(\Dsa \to D^{*+}K^0_S)}{{\cal B}(\Dsa \to D^+K^0_S)
} = \\
 1.14 \pm 0.39_{\sta} \pm 0.23_{\sys}
\end{split}
\end{equation}

\begin{equation}
\begin{split}
r_4 = \frac{
  {\cal B}(\Dsb \to D^{*+}K^0_S)}{{\cal B}(\Dsb \to D^+K^0_S)
} = \\
 1.38 \pm 0.35_{\sta} \pm 0.49_{\sys}
\end{split}
\end{equation}

where $D^{*+} \to D^+ \pi^0$.

Averaging $r_1$, $r_3$ and $r_2$, $r_4$ we obtain:

\begin{equation}
\begin{split}
\frac{{\cal B}(\Dsa \to D^*K)}{{\cal B}(\Dsa \to D K)} = 
0.91 \pm  0.13_{\sta} \pm  0.12_{\sys}
\end{split}
\end{equation}

\begin{equation}
\begin{split}
\frac{{\cal B}(\Dsb \to D^*K)}{{\cal B}(\Dsb \to D K)} = 
1.10 \pm 0.15_{\sta} \pm 0.19_{\sys}.
\end{split}
\end{equation}

We also make a test of isospin conservation. We use the yields obtained from the fit to the appropriate mass spectra
and correct for efficiency and branching fractions. We obtain, within the errors, similar rates for resonance 
decays to $D^0 K^+$ and  $D^+ \KS$ as well as for decays to $D^{*0} K^+$ and  $D^{*+} \KS$, as expected from isospin
conservation.

We now compare these results with theoretical expectations.
In the work of Ref.~\cite{colangelo2}, for $J^P=1^-$, two different quark model assignments are proposed for the $\Dsa$:
the $l=2$ ground state, $1^3D_1$, and the $l=0$ first radial excitation, $2^3S_1$. In the first case the ratio 
${\cal B}(D^*_{s1}(2710) \to D^*K)/{\cal B}(D^*_{s1}(2710) \to D K)$ is expected to be $0.043\pm0.002$, in the 
second case $0.91\pm0.04$. In this framework the $\Dsa$ can be identified as the first radial excitation of the $D_s^*(2112)$.
The same assignment is derived from Ref.~\cite{close} where a mass of 2711 \mevcc is predicted 
for the state $2^3S_1$. However, in this case the expected ratio is 3.55, in significant disagreement with the measured value.

In Ref.~\cite{nowak}, in the framework of chiral doublers, $J^P=1^-$ states are expected at masses of 2632 and 2720 \mevcc.

The observation of $\Dsb \to D^*K$ rules out the $J^{P}=0^+$ assignment suggested by Refs.~\cite{port,close}.
In Ref.~\cite{colangelo1} the $J^P=3^-$ assignment is proposed; however the
predicted ${\cal B}(D^*_{sJ}(2860) \to D^*K)/{\cal B}(D^*_{sJ}(2860) \to D K)$ is
0.39, which differs from our measurement at
the level of three standard deviations. A better agreement is obtained if we compare with the calculations from Ref.~\cite{Liu}
which expects a ratio of 0.6.

As to the possible interpretation of the $\Dsc$ state, we note that
Ref.~\cite{matsuki} predicts two $J^{P}=1^+$ radial excitations at 3082 and 3094 \mevcc.

\section{\boldmath Conclusions} 
 
In summary, in 470~${\rm fb}^{-1}$ of data collected
by the \babar\  experiment, 
we observe the decays of the $\Dsa$ and $\Dsb$ to $D^*K$ and measure their branching fractions
relative to $D K$. A new, broad $D_{sJ}^+$ state is observed in the $D^*K$ mass spectrum at a mass near 3040 \gevcc.
Possible spin-parity assignments for these states are discussed. 

\section{\boldmath Acknowledgments}
We are grateful for the 
extraordinary contributions of our \pep2\ colleagues in
achieving the excellent luminosity and machine conditions
that have made this work possible.
The success of this project also relies critically on the 
expertise and dedication of the computing organizations that 
support \babar.
The collaborating institutions wish to thank 
SLAC for its support and the kind hospitality extended to them. 
This work is supported by the
US Department of Energy
and National Science Foundation, the
Natural Sciences and Engineering Research Council (Canada),
the Commissariat \`a l'Energie Atomique and
Institut National de Physique Nucl\'eaire et de Physique des Particules
(France), the
Bundesministerium f\"ur Bildung und Forschung and
Deutsche Forschungsgemeinschaft
(Germany), the
Istituto Nazionale di Fisica Nucleare (Italy),
the Foundation for Fundamental Research on Matter (The Netherlands),
the Research Council of Norway, the
Ministry of Education and Science of the Russian Federation, 
Ministerio de Educaci\'on y Ciencia (Spain), and the
Science and Technology Facilities Council (United Kingdom).
Individuals have received support from 
the Marie-Curie IEF program (European Union) and
the A. P. Sloan Foundation.


\begin{thebibliography}{99}
\bibitem{footnote}
The use of charge-conjugate reactions is implied throughout.
\end{thebibliography}
\end{document}